\documentstyle[11pt,newpasp,twoside,epsf]{article}
\def\arcsec{\ifmmode '' \else $''$\fi}
\def\arcmin{\ifmmode ' \else $'$\fi}
\def\arcsecpoint{\ifmmode ''\!. \else $''\!.$\fi}
\def\arcminpoint{\ifmmode '\!. \else $'\!.$\fi}
\def\kms{\ifmmode {\rm km\ s}^{-1} \else km s$^{-1}$\fi}
\def\Hubble{\ifmmode {\rm km\ s}^{-1}\ {\rm Mpc}^{-1} 
	\else km s$^{-1}$ Mpc$^{-1}$\fi}
\def\ergsec{\ifmmode {\rm ergs\ s}^{-1} \else ergs s$^{-1}$\fi}
\def\eflux{\ifmmode {\rm ergs\ s}^{-1}\;{\rm cm}^{-2}
	  \else ergs s$^{-1}$ cm$^{-2}$\fi}
\def\efluxA{\ifmmode {\rm ergs\ s}^{-1}\;{\rm cm}^{-2}\;{\rm \AA}^{-1}
	  \else ergs s$^{-1}$ cm$^{-2}$ \AA$^{-1}$\fi}
\def\efluxHz{\ifmmode {\rm ergs\ s}^{-1}\;{\rm cm}^{-2}\;{\rm Hz}^{-1}
	  \else ergs s$^{-1}$ cm$^{-2}$ Hz$^{-1}$\fi}
\def\cc{\ifmmode {\rm cm}^{-3} \else cm$^{-3}$\fi}
\def\vFWHM{\ifmmode v_{\mbox{\tiny FWHM}} \else
            $v_{\mbox{\tiny FWHM}}$\fi}
\def\Msun{\ifmmode M_{\odot} \else $M_{\odot}$\fi}
\def\Lsun{\ifmmode L_{\odot} \else $L_{\odot}$\fi}
\def\Halpha{\ifmmode {\rm H}\alpha \else H$\alpha$\fi}
\def\Hbeta{\ifmmode {\rm H}\beta \else H$\beta$\fi}
\def\Hgamma{\ifmmode {\rm H}\gamma \else H$\gamma$\fi}
\def\Hdelta{\ifmmode {\rm H}\delta \else H$\delta$\fi}
\def\Lya{\ifmmode {\rm Ly}\alpha \else Ly$\alpha$\fi}
\def\Lyb{\ifmmode {\rm Ly}\beta \else Ly$\beta$\fi}
\def\hi{\ifmmode {\rm H}\,{\sc i} \else H\,{\sc i}\fi}
\def\hii{\ifmmode {\rm H}\,{\sc ii} \else H\,{\sc ii}\fi}

\def\ciii{\ifmmode {\rm C}\,{\sc iii} \else C\,{\sc iii}\fi}
\def\civ{\ifmmode {\rm C}\,{\sc iv} \else C\,{\sc iv}\fi}

\begin{document}
\title{The polarization properties of Broad Absorption Line QSOs: observational results}
\author{D.~Hutsem\'ekers\footnotemark[1]}
\footnotetext[1]{Also Research Associate FNRS, University of Li\`ege, Belgium}
\affil{European Southern Observatory, Chile}
\author{H.~Lamy}
\affil{University of Li\`ege, Belgium}
\begin{abstract}
Correlations between BAL QSO intrinsic properties and polarization
have been searched for. Some results are summarized here, providing
possible constraints on BAL outflow models.
\end{abstract}
\section{Introduction}
From 1994 to 1999 we have obtained broad-band linear polarization
measurements for a sample of approximately 50 Broad Absorption Line
(BAL) QSOs using the ESO 3.6m telescope at La Silla (Chile).\\
On the basis of this sample plus additional data compiled from the
literature, possible correlations between BAL QSO intrinsic properties
and polarization have been searched for. Here we present some of our
most interesting results, updated with recent data.
\section{Analysis and results}
A careful distinction between BAL QSO subtypes has been done. In
addition to the BAL QSOs with high-ionization (HI) absorption features
only, we have distinguished BAL QSOs with strong (S), weak (W), and
marginal (M) low-ionization (LI) absorption troughs (Hutsem\'ekers et
al. 1998, 2000 for details).\\
Several indices are used to quantify the spectral characteristics: the
balnicity index (BI) which is a modified velocity equivalent width of
the $\civ$ BAL, the detachment index (DI) which measures the degree of
detachment of the absorption trough relative to the emission line, the
maximum velocity ${\sc\sl v}_{\rm max}$ in the $\civ$ BAL, and the
power-law index $\alpha$ of the continuum.\\
Although most BAL QSOs are radio-quiet, some of them appear
radio-moderate, and radio-to-optical flux ratios $R^{\star}$ were also
collected.\\
Correlations and sample differences were searched for by means of the
usual statistical tests. Survival analysis was used for censored data
(mainly $R^{\star}$). While the study of polarization was our main
goal, correlations between different indices have also been considered.\\
Results presented by Hutsem\'ekers et al.~(1998, 2000; H1998, H2000)
are updated with polarimetric data from Schmidt \& Hines (1999;
S1999), Lamy \& Hutsem\'ekers (2000; L2000), and Ogle et al.~(2000;
O2000). Only polarimetric measurements with $\sigma_p \leq 0.4\%$ are
taken into account, such that the debiased polarization degree $p_0$
has a typical uncertainty $\sigma_p$ = 0.2-0.3\%.  The radio-loud BAL
QSOs recently discovered in the FIRST survey (Becker et al. 2000;
B2000) are included in the present study.\\[0.5cm]
$\bullet$ {\bf Evidence for polarization differences between low- and high-ionization BAL QSOs}\\
\begin{figure}%
\leavevmode \centering
\epsfysize=7.0cm \epsfbox{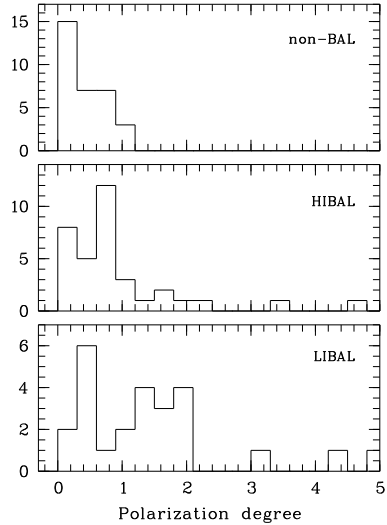}
\caption{The distribution of the polarization degree $p_0$ (in \%) for
the three main classes of QSOs. LIBAL QSOs contain the three
sub-categories, i.e. strong, weak and marginal LIBAL QSOs.  Data are
from H1998, S1999, L2000, O2000 (a LIBAL QSO with $p_0$=7.5 and the
unclassified BAL QSOs are not represented here)}
\end{figure}%
The distribution of the polarization degree $p_0$ for the three main
classes of QSOs is illustrated in Fig.~1.  We can see that the bulk of
QSOs with $p_0 >$ 1.2\% belong to the sub-class of LIBAL QSOs.  Note
that not all LIBAL QSOs are highly polarized.  As a class, HIBAL QSOs
appear less polarized than LIBAL QSOs and more polarized that non-BAL
QSOs. They seem to have intermediate properties.  All these
differences are statistically significant ($P_{\rm\sc k-s} \geq$
99\%).\\[0.3cm]
$\bullet$ {\bf The correlation between the balnicity and the slope of the continuum}\\
In addition to their higher polarization, it is seen from Fig.~2 that
most LIBAL QSOs have also larger balnicities and more reddened
continua than HIBAL QSOs.  Considering the whole BAL QSO sample
(i.e. HI+LI BALs), a significant ($P_{\tau}\geq$ 99\%) correlation is found
between the balnicity index BI and the slope of the continuum. Since
LIBAL QSOs as a class are more reddened and more polarized than HIBAL
QSOs, it also results a correlation between the power-law index and the
polarization, although less convincing.\\[0.3cm]
$\bullet$ {\bf The  correlation between the polarization of the continuum and the line profile detachment index}\\
Among several possible correlations of the polarization with spectral
indices like the balnicity index, the equivalent width and the velocity
width of $\civ$ and $\ciii${\small ]}, the only significant ($P_{\tau}
\geq$ 99\%) correlation we found is a correlation with the line
profile detachment index, quite unexpectedly. Fig.~2 illustrates the
correlation between the polarization degree $p_0$ and the line profile
detachment index DI for all BAL QSOs of our sample.  The correlation
is especially apparent and significant for the LIBAL QSOs. It
indicates that the BAL QSOs with P Cygni-type line profiles (DI$\ll$)
are the most polarized.\\[0.3cm]
\begin{figure}%
\leavevmode 
\epsfysize=6.15cm \epsfbox[1 1 531 521]{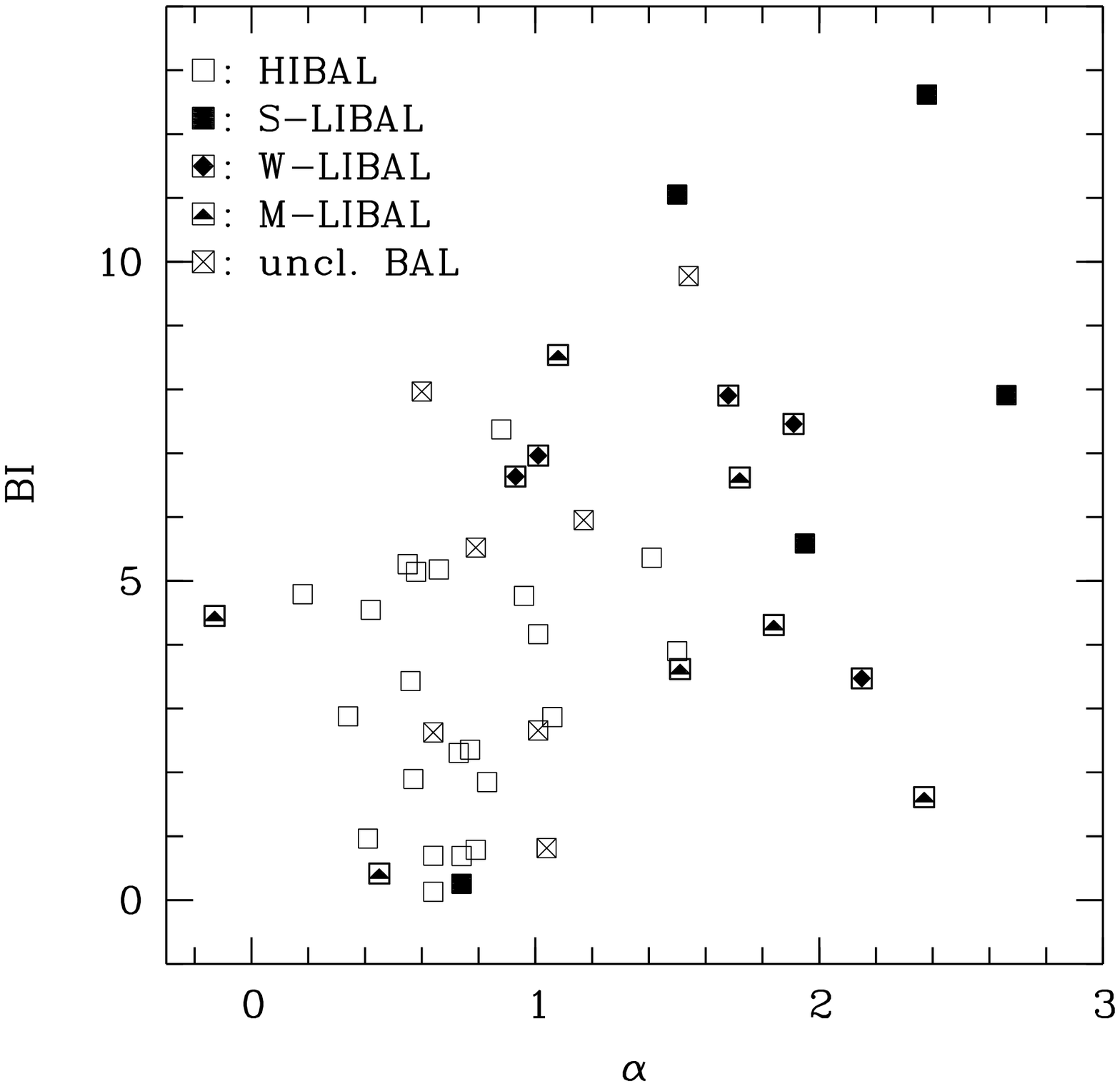}  \hspace*{0.4cm}
\epsfysize=6.15cm \epsfbox[5 5 535 521]{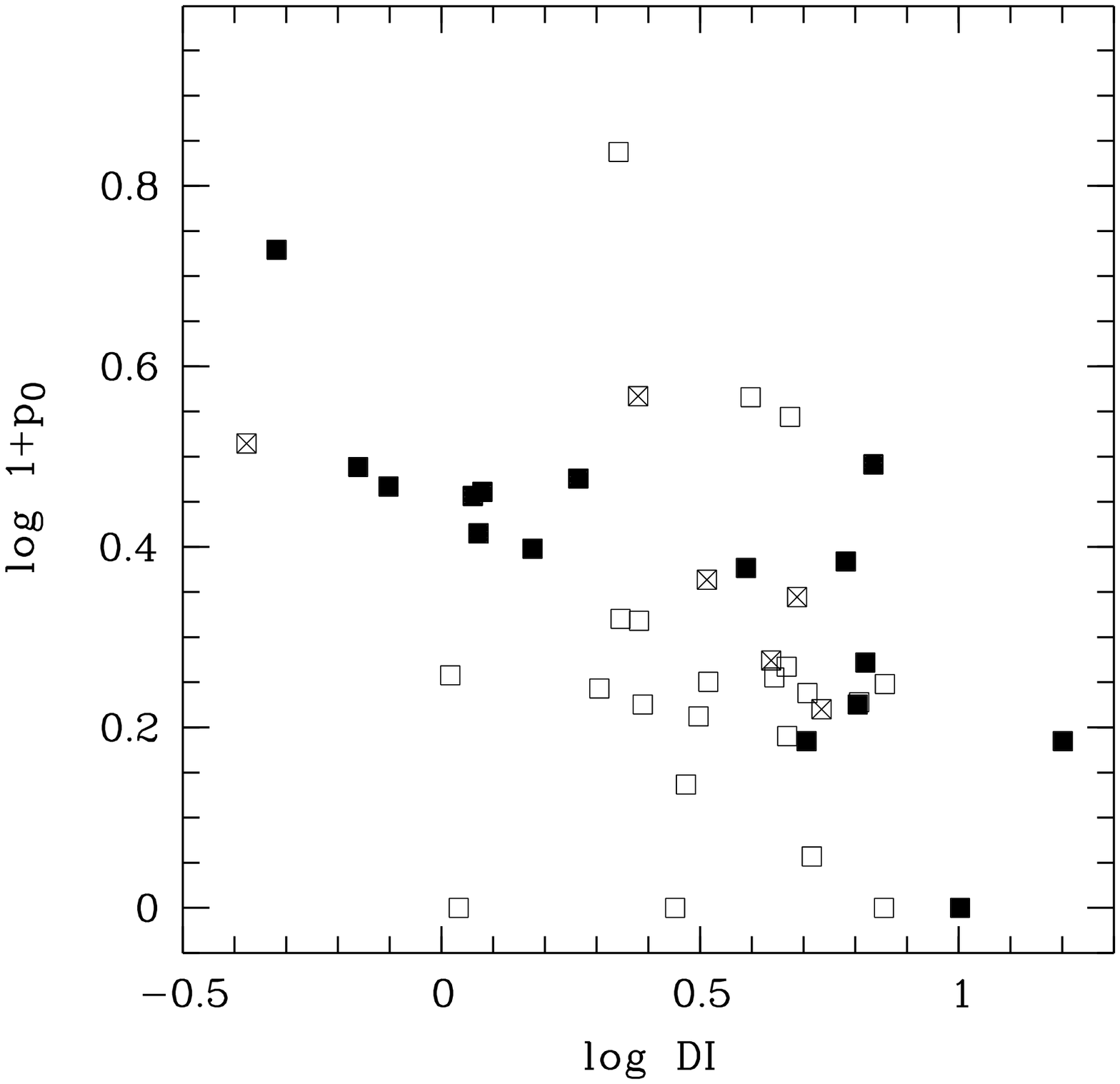}
\caption{{\bf Left}: the correlation between the balnicity index BI
(in 10$^{3}$ km s$^{-1}$) and the power-law index $\alpha$ ($F_{\nu}
\propto \nu^{-\alpha}$).  The 3 sub-categories of LIBAL QSOs are
distinguished here.  Data and objects are from H1998, L2000,
H2000. {\bf Right}: the correlation between the polarization degree
$p_0$ (in \%) and the line profile detachment index DI. The
correlation is especially apparent for the LIBAL QSOs (filled
squares). Data from H1998, S1999, L2000}
\end{figure}%
$\bullet$ {\bf The absence of correlation between the polarization and $R^{\star}$}\\
If the higher polarization of BAL QSOs as a class is due to an
attenuation of the direct continuum with respect to the scattered one
--at least in some objects-- (Goodrich 1997), then we expect the
polarization to be correlated with the radio-to-optical flux ratio.
In Fig.~3, the BAL QSO polarization $p_0$ is plotted against the
radio-to-optical flux ratio $R^{\star}$. No correlation is seen, as
confirmed by the statistical tests.  Note that the distribution of
$R^{\star}$ is not found to differ between the HIBAL and LIBAL
subsamples\\[0.3cm]
$\bullet$ {\bf The absence of correlation between the terminal velocity and $R^{\star}$}\\
In order to investigate the claimed anticorrelation between the
terminal velocity of the flow and the radio-to-optical flux ratio
(Weymann 1997), we have plotted in Fig.~3 the maximum velocity
${\sc\sl v}_{\rm max}$ in the $\civ$ BAL against the radio-to-optical
flux ratio $R^{\star}$.  No correlation is found, as confirmed by the
statistical tests. 
\begin{figure}%
\leavevmode 
\epsfysize=6.2cm \epsfbox[1 4 531 511]{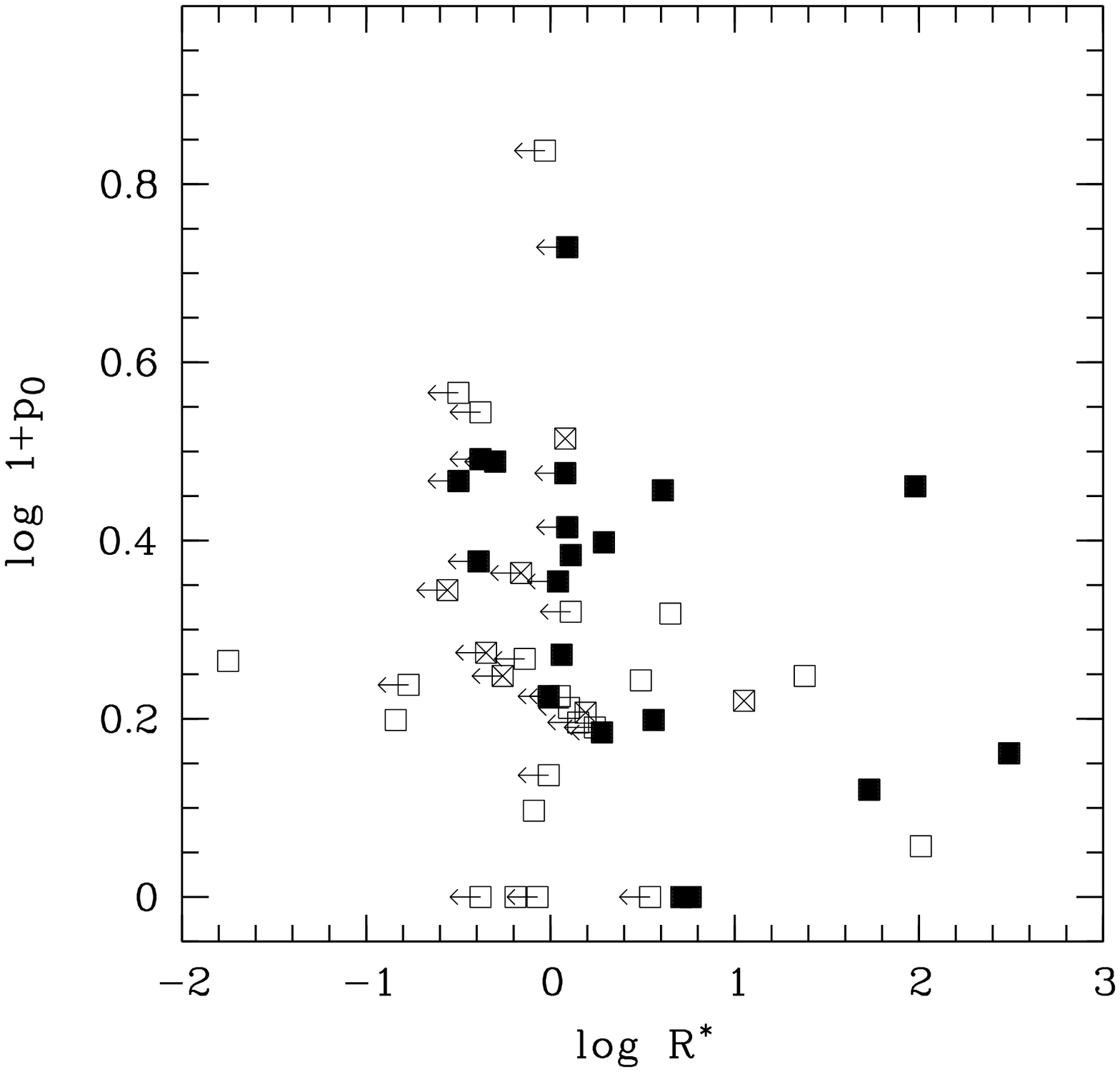} \hspace*{0.3cm}
\epsfysize=6.2cm \epsfbox[4 4 522 513]{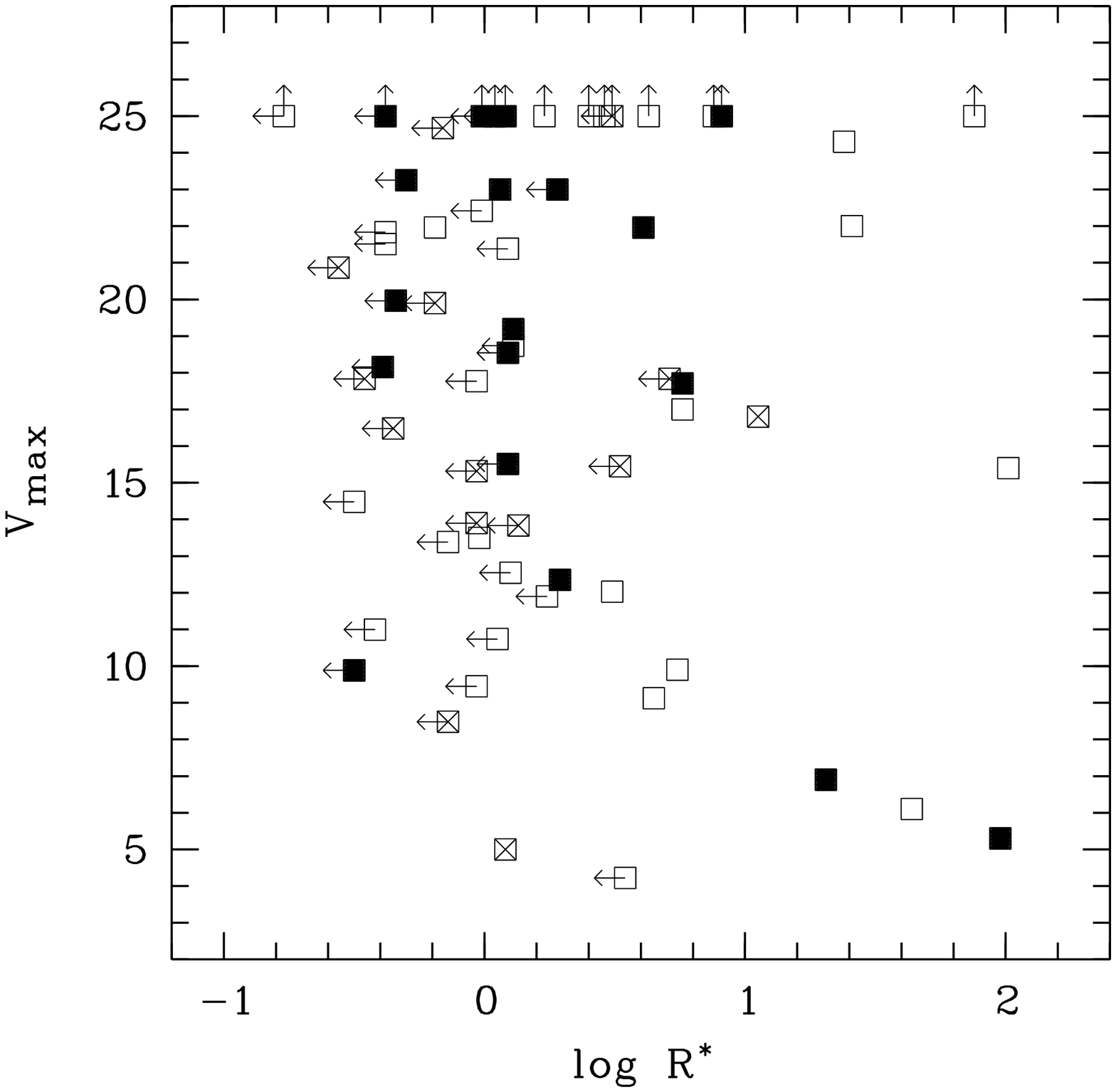}
\caption{{\bf Left}: The polarization degree $p_0$ plotted against the
radio-to-optical flux ratio $R^{\star}$.  Data are from H2000. {\bf
Right}: the maximum velocity ${\sc\sl v}_{\rm max}$ in the $\civ$ BAL
(in 10$^{3}$ km s$^{-1}$) is plotted against the radio-to-optical flux
ratio $R^{\star}$. Data from H2000, B2000. In both figures, open
squares represent HIBAL QSOs, filled squares LIBAL QSOs, and squares
with a cross unclassified BAL QSOs, while arrows indicate censored
data
points}
\end{figure}%
\section{Conclusions}
Our results show that polarization is correlated with BAL QSO line
profiles and types, emphasizing the extreme behavior of LIBAL QSOs
already reported by several studies. These results could provide
constraints on the BAL outflow models and geometry. As discussed by
Hutsem\'ekers et al. (1998), they are consistent with the Murray et
al.~(1995) disk-wind model.
\end{document}